# Structural, Electronic, and Dynamical Properties of Liquid Water by *ab initio* Molecular Dynamics based on SCAN Functional within the Canonical Ensemble

Lixin Zheng[1,a], Mohan Chen[1,a], Zhaoru Sun[1], Hsin-Yu Ko[2], Biswajit Santra[2], Pratikkumar Dhuvad[1], and Xifan Wu[1,3]

[1]*Department of Physics, Temple University, Philadelphia, PA 19122 USA*

[2]*Department of Chemistry, Princeton University, Princeton, New Jersey 08544, USA*

[3]*Institute for Computational Molecular Science, Temple University, Philadelphia, PA 19122, USA*

## ABSTRACT

We perform *ab initio* molecular dynamics (AIMD) simulation of liquid water in the canonical ensemble at ambient conditions using the SCAN meta-GGA functional approximation, and carry out systematic comparisons with the results obtained from the GGA-level PBE functional, and Tkatchenko-Scheffler van der Waals (vdW) dispersion correction inclusive PBE functional. We analyze various properties of liquid water including radial distribution functions, oxygen-oxygen-oxygen triplet angular distribution, tetrahedrality, hydrogen bonds, diffusion coefficients, ring statistics, density of states, band gaps, and dipole moments. We find that the SCAN functional is generally more accurate than the other two functionals for liquid water by not only capturing the intermediate-range vdW interactions but also mitigating the overly strong hydrogen bonds prescribed in PBE simulations. We also compare the results of SCAN-based AIMD simulations in the canonical and isothermal-isobaric ensembles. Our results suggest that SCAN provides a reliable description for most structural, electronic, and dynamical properties in liquid water.

---

[a] Electronic mail: lixin.zheng@temple.edu, mohan.chen@temple.edu

# I. INTRODUCTION

Liquid water is one of the most important solvents for chemical and biological processes [1]. Bonded by relatively strong covalent bonds, the structure of a single water molecule is now well understood. However, in its condensed phases, water molecules form tetrahedral or near–tetrahedral structures through hydrogen bonds (H-bonds). Within such a network, water molecules are bonded to each other through directional H-bonds resulting from weak electrostatic attractions between a lone pair electron of oxygen (O) and a hydrogen (H) from the neighboring water molecule. In contrast to covalent bonds, the H-bonds are energetically much weaker. Therefore, the structure of the H-bond network is not only easily disturbed by thermal fluctuations and applied pressures as evidenced by the richness of ice phases and liquid structures [2-4] but also largely affected by relatively weak physical interactions such as van der Waals (vdW) dispersion forces [5-7]. Furthermore, the above picture becomes even more complex when the quantum nature of the nuclei is taken into consideration [8-10]. Because of the delicacy of the H-bond network, the structure of liquid water has neither been concluded from experiment nor from theory. Therefore, the nature of the H-bond structure in liquid water continues to be at the center of scientific debate. Not surprisingly, research in this field is progressing by continuous joint efforts from both experiment and theory.

Several experimental techniques were used to probe the microscopic properties of liquid water, including X-ray and neutron scattering [11], X-ray absorption spectroscopy (XAS) [12], infrared spectroscopy [13], Raman spectroscopy [14], and photoemission spectroscopy (PES) [15,16], etc. X-ray and neutron scattering experiments directly measure the statistically averaged structural information. On the one hand, X-ray is more sensitive to O than H atoms since it is scattered by electrons and can only capture O-O correlations with high clarity. On the other hand, neutrons scattered from nuclei are ideal for obtaining correlations involving H atoms. The scattered intensities are used to extract the O-O, O-H, and H-H radial distribution functions (RDFs). However, experimental determination of the RDFs is very challenging and typically requires scattering experiments with isotope substitutions and theoretical input such as empirical potential structural refinement (EPSR) [10-12,17]. In addition, XAS measurements are able to probe the electronic structure in different phases of water, from which the tetrahedral H-bonding and shell structure information are indirectly obtained [12]. Furthermore, the infrared and Raman spectroscopies allow the extraction of the O-H vibrational dynamics of water [13,14]. Finally, the valence band photoemission experiments measure the occupied electronic states, which reveal the local bonding environment in liquid water [15,16].

Nevertheless, due to breaking and reforming of H-bonds on the timescale of picoseconds (ps) [18], an experimental technique that could directly detect the real-space microscopic structure of liquid water is still lacking. Importantly, structural properties of liquid water could be obtained via computer simulations. For example, *ab initio* molecular dynamics (AIMD) [19] simulations employing density functional theory (DFT) [20] model a given system based on quantum mechanical principles and in general needs no empirical inputs. Specifically, AIMD simulations generate a nuclear potential energy surface "on the fly" from the electronic ground state and yield a variety of properties including structural, electronic, and dynamical properties of the system. Therefore, AIMD is an ideal approach for modeling important properties of water across the phase diagram [21-24]. Indeed, during the past two decades, much progress has been made on the AIMD simulations of liquid water by adopting different levels of functional approximations [25-33]. These works have advanced our understandings on various properties of liquid water and the ice phases. Other methods have been also applied to liquid water in recent years, e.g. Monte Carlo simulations [34,35], Møller-Plesset perturbation theory (MP2) [36], and random phase approximation [37]. However, DFT is still the most popular ab initio method to study liquid water due to the balance between accuracy and computational cost.

However, challenges still remain in providing a more accurate water model by DFT-based AIMD simulations, which depend crucially upon the underlying exchange-correlation (XC) functional. For instance, gas-phase water clusters were first modeled by the local density approximation (LDA) [38] with apparently over-estimated H-bond strengths and distances between water molecules that are too close. Currently, one of the most widely used category of XC functionals for AIMD simulations of liquid water is the generalized-gradient approximation (GGA), which largely corrects the over-binding provided by the LDA functional but still largely deviate from the experiments. Among various GGA-level XC functionals, the Perdew-Burke-Ernzerhof (PBE) [39] and Becke-Lee-Yang-Parr (BLYP) [40,41] functionals are two of the most popular forms [25-27,29,42]. However, the modeling of liquid water by GGA still exhibits a rigid solvation shell structure and much smaller diffusivity when compared to experimental data. Furthermore, the inherent self-interaction error (SIE) existing in the GGA-level functionals unavoidably leads to excessive proton delocalization in liquid water and overestimation of the molecular polarizability [43,44]. What's worse, GGA functionals fail to describe the correct density ordering between liquid water and hexagonal ice, leading to the consequence that the GGA ice would sink in GGA water [5,27,42]. The above is due to the fact that GGA ignores the non-local electron correlation effects that are responsible for vdW or dispersion interactions, which are crucial for improving water density properties [5,27,45,46]. Although the short-range portion of the vdW interactions is described by the local and semi-local XC functionals, the intermediate- and long-range parts are not captured by any general-purpose GGA functionals. Besides the choice of XC functional, the nuclear quantum effects (NQEs) are essential for light atoms such as hydrogen and deuterium, which are not considered in the framework of classical molecular dynamics. Studies [8,9] have shown that the inclusion of NQEs in AIMD simulations further softens the shell structure of water molecules in the liquid phase. Therefore, an accurate description of the balance among covalent bonds, H-bonds, and vdW interactions in liquid water that could eventually improve its H-bond network remains a challenge.

Several studies have tried to address the issues in different ways. For instance, to reconcile the absence of intermediate- and long-range vdW interactions, several empirical [47] and non-empirical [48] vdW corrections have been proposed. Particularly, the vdW density functional [5,49] and density-dependent Tkatchenko-Scheffler dispersion correction [6,27,28,50] were found to soften the solvation shell structures of water molecules and largely correct the density of liquid water. Moreover, hybrid XC functional such as PBE0, wherein a fraction of the exact exchange energy is included in the density functional approximation, was used for the simulation of liquid water in order to mitigate the SIE [27,28,51]. Properties of liquid water with the use of the PBE0 and Tkatchenko-Scheffler dispersion correction were thoroughly studied in Ref. [28], and it was found that the substitution of PBE by PBE0 functional in the canonical ensemble has similar effects as vdW interactions which soften the shell structures of water molecules. However, by only including PBE0, the density of liquid water is further lower compared to that from PBE [27]. In summary, although simulations of liquid water utilizing GGA XC functionals exhibited several critical shortages on the description of the H-bond network in liquid water, other non-empirical, general-purpose XC functionals that could describe all types of interactions in liquid water on an equal footing are still largely absent. Instead, one could only combine different XC functionals, most of which include empirical parameters, to improve the description of liquid water at the molecular level.

The recently developed strongly constrained and appropriately normed (SCAN) meta-GGA functional [52] can be used within AIMD approach to address the above issues. SCAN satisfies all 17 exact constraints appropriate for semilocal XC functionals and has been shown to predict accurate geometries and energies of diversely bonded molecules and condensed matter materials while maintaining the computational cost similar to GGA [52,53]. Furthermore, Ref. [53] reports that SCAN functional is highly accurate for covalent bonds in water monomers, it also captures the intermediate-range vdW interactions in ice and gas-phase water hexamers. This study implies that SCAN possesses the ingredients necessary to describe liquid water. In a recent work [54], AIMD simulations of liquid water in the isothermal-isobaric (NpT) ensemble employing the SCAN functional were carried out, reporting that SCAN correctly predicts higher density of liquid water than ice and gives a delicately balanced description of covalent bonds, H-bonds, and vdW interactions in liquid water.

In this work, in order to compare the different vdW effects in the SCAN and PBE+vdW functionals, as well as the effects of different ensembles on properties of liquid water, we performed AIMD simulations of liquid water in the canonical (NVT) ensemble with PBE, PBE+vdW, and SCAN functionals. We demonstrate in detail that a variety of properties of liquid water by SCAN are systematically improved towards the experimental direction when compared to the semi-local PBE and PBE+vdW functionals. Essentially, SCAN improves the description of electronic structure of liquid water and captures the intermediate-range vdW interactions. As a result, structural properties including radial distribution functions, local tetrahedrality, oxygen-oxygen-oxygen angular distribution function, and statistics of H-bond network properties, as well as electrostatic properties such as molecular dipole moment and band gap in liquid water have been systematically improved. We find that both SCAN and PBE+vdW functionals outperform the PBE functional. Moreover, although vdW interactions can be described by both SCAN and PBE+vdW functionals, the two functionals perform differently; SCAN has a more significant improvement over the structural properties of liquid water than PBE+vdW does. This is understood by the fact that the vdW functional does not affect the directional H-bond strengths, while SCAN improves the covalent bonds in water molecules and reduces the overly strong directional H-bond strength as described by the PBE functional. Besides, we also compare the simulation results between the NVT and the NpT ensembles, which illuminate how ensembles and slightly different water densities influence the structural properties of liquid water.

The paper is organized as follows. First, we describe the simulation methods and list the computational details in Section II. We then show the AIMD simulations results of structural, electronic, and dynamical properties of liquid water as obtained by employing different XC functionals in Section III. Finally, we conclude the results in Section IV.

## II. METHODS

All simulations were performed with the Car-Parrinello (CP) molecular dynamics [19] implemented in the Quantum ESPRESSO package [55,56]. We set up a cubic cell with $L$=12.445 Å and periodic boundary conditions which consists of 64 $H_2O$ molecules. The electronic wavefunctions were expanded using a plane wave basis set with an energy cutoff of 85 Ry. The norm-conserving pseudopotentials [57] in the form of Hamann-Schlüter-Chiang-Vanderbilt (HSCV) were adopted for the O and H atoms. The functionals employed herein include the semi-local PBE GGA functional, PBE with the non-local vdW/dispersion interactions in the form of the density-dependent Tkatchenko-Scheffler [50] dispersion correction, i.e., PBE+vdW, and the SCAN meta-GGA functional. The parameters used for the Tkatchenko-Scheffler vdW were adopted from Ref. [28].

All AIMD simulations were performed in the NVT ensemble at 330 K by following Ref. [28]. The PBE, PBE+vdW, and SCAN systems were initially equilibrated for about 8 ps, and then continued for 55, 54, and 50 ps, respectively, for data collection. The nuclear mass of deuterium, 2.014 amu, was used for each hydrogen atom, and the mass of oxygen is 15.999 amu. The ionic temperature was controlled by a single Nosé-Hoover thermostat [58-60] with a frequency of 60 THz. The CP equations of motion were integrated using the standard Verlet algorithm with a time step of 2.0 a.u. (~0.0484 fs). To ensure an adiabatic separation between the nuclear and electronic degrees of freedom in the CP dynamics, an appropriate electronic fictitious mass $\mu$ should be chosen in CP molecular dynamics. It was discussed in Refs. [26] that compared to Born-Oppenheimer (BO) dynamics, simulations of liquid water with CP dynamics yield lower maximums and higher minimums in the radial distribution functions. In particular, a larger $\mu$ used in CP dynamics further softens the structure of liquid water. In this work, we set the electronic fictitious mass $\mu$ to be 100 a.u. with an electronic mass cutoff of 25 a.u. [61]. Additionally, we compare the NVT data to those obtained with CP molecular dynamics simulations employing PBE and SCAN functionals in the NpT ensemble, and the parameters of NpT simulations are described in Ref. [54].

## III. RESULTS AND DISCUSSION

### A. Comparative analysis of radial distribution functions

We compare the computed results of radial distribution functions with the experimental results. In each of the O-O, O-H, and H-H radial distribution functions, we utilize the experimental data obtained from the EPSR based on joint X-ray/neutron data by Soper and Benmore [10] and compare them with our simulation results performed with PBE, PBE+vdW, and SCAN functionals. In general, we find that while both PBE+vdW and SCAN functionals soften the structure of liquid water compared to that obtained from the PBE functional, SCAN has a larger correction than PBE+vdW does and the reasons are explicated in the following paragraphs.

Fig. 1 shows the O-O radial distribution function, $g_{OO}(r)$, of liquid water. The intensities and positions of maximums and minimums of $g_{OO}(r)$ found from our simulations, together with those measured by the X-ray diffraction experiments with high Q-range and low statistical noise [17], are shown in Table I. At the PBE level of theory, due to the lack of intermediate- and long-range vdW interactions and the SIE that lead to overly strong H-bonds, the liquid water system has a relatively more rigid shell structure compared to experiment. By including vdW correction in the PBE simulation, the structure of liquid water is improved, which is consistent with previous reports [5,27,28,45,51]. The $g_{OO}(r)$ is softened because of the non-directional vdW interactions which provide attractive forces between a given molecule and the molecules around it. Therefore, the water molecules originally located at the second coordination shells move inwards to the interstitial region between the first and second shells. Meanwhile, some water molecules in the first shell are pushed outwards due to the breaking of H-bonds. As a result, the locally disordered configurations with O-O distance in the interstitial region are energetically more favorable. In other words, the inclusion of vdW interactions corrects the overly-strong H-bonding network by populating the interstitial region and softening the over-structured configurations. However, since vdW interactions alone do not change the description of electronic structure of liquid water, the overly strong directional H-bonding strength prescribed in the PBE density functional remains unchanged.

By using the SCAN XC functional, we observe that the $g_{OO}(r)$ of liquid water is further softened as compared to the $g_{OO}(r)$ from the PBE+vdW AIMD simulation. Specifically, as shown in both Fig. 1 and Table I, the intensities of the first maximum ($g_1^{max}$), first minimum ($g_1^{min}$), and second maximum ($g_2^{max}$) are all improved; additionally, SCAN corrects the position of the first minimum, $r_1^{max}$ towards the experimental value. The excellent performance by SCAN to soften the structure of liquid water is partly attributed to its ability to capture the intermediate-range vdW interactions. Moreover, SCAN provides improvement over the short-range interactions in liquid water, i.e., the O-H covalent bond and the overly strong directional H-bonds as described by PBE. As will be discussed with more details in the next paragraph and Sec III E, we find that SCAN improves the length of O-H covalent bond and electronic structure of liquid water. Consequently, the weakened H-bonds lead to more broken H-bonds and more disordered local tetrahedral structure in liquid water, which contribute to the additional softening of the shell structure observed in $g_{OO}(r)$. Although the $g_{OO}(r)$ obtained from SCAN and the experimental result are almost identical beyond 3.5 Å, we notice that within the first coordination shell SCAN still gives slightly over-structured liquid water as compared to the experimental result. The still over-structured $g_{OO}(r)$ described by SCAN in the first coordination shell is possibly due to the lack of NQEs and the existence of SIE in our current modeling[8,28,43], and can be taken into account with more advanced simulation methods such as hybrid SCAN functional and will be addressed in future work.

We now turn our attention to the O-H radial distribution function, $g_{OH}(r)$, as shown in Fig. 2, where a similar trend of systematic softening of structure can be observed with the use of PBE+vdW and SCAN functionals as compared to the PBE results. Again, we find that SCAN outperforms PBE+vdW for the description of the $g_{OH}(r)$ in several ways. First, the position of the first maximum of the $g_{OH}(r)$, $r_1^{max}$, which corresponds to the O-H covalent bond, is shortened from 0.994 Å in PBE to 0.978 Å in SCAN AIMD simulations; however, there is only a minor change in the PBE+vdW simulation, which gives the O-H bond length of 0.990 Å. This shorter length of covalent bond as described by SCAN confirms the enhancement of the covalency of water molecules. This result is also consistent with the previous work [53], where it was reported that SCAN is highly accurate for properties of gas-phase water monomers including O-H covalent bond length and H-O-H covalent bond angle. As the O and H atoms in a molecule bind more strongly, it is more difficult for an H atom to form a H-bond with its neighboring O atoms, resulting in more broken H-bonds in a softer structure of liquid water.

The second maximum of the $g_{OH}(r)$ helps to understand better the H-bond network in liquid water, as it corresponds to the intermolecular O-H distances. Table II shows that both PBE+vdW and SCAN yield intensities of the second maximum $g_2^{max}$ of the $g_{OH}(r)$ closer to the experimental value as compared to PBE. More interestingly, we observe that SCAN outperforms PBE+vdW in correcting both $g_2^{max}$ and $g_2^{min}$ (the second minimum) of the $g_{OH}(r)$ because the two functionals disrupt the H-bond network of liquid water in slightly different ways. On one hand, in the PBE+vdW simulation, the interstitial region between the first and second solvation shells is populated due to the attractive forces among water molecules provided by vdW interactions, which further disrupts the first shell; however, strengths of directional H-bonds in liquid water are barely changed. On the other hand, SCAN not only captures the intermediate-range vdW interactions but also corrects the unphysical overly strong directional H-bonding strength prescribed in PBE; both corrections by SCAN contribute to the improved description of the H-bond network in liquid water. We notice that both PBE+vdW and SCAN functionals increase the position of the second maximum, $r_2^{max}$, from 1.72 Å in PBE to 1.75 Å in PBE+vdW, and 1.79 Å in SCAN, while the experimental value is 1.73 Å. Finally, for the third maximum of the $g_{OH}(r)$, we observe that the intensity of $g_3^{max}$ from PBE+vdW only reduces by 0.01 with a 0.02 Å shift of the position $r_3^{max}$ outwards as compared to PBE, while SCAN reduces $g_3^{max}$ by 0.08 and a 0.04 Å shift of $r_3^{max}$ outwards compared to PBE. As a result, the difference between the intensity of the third maximum and the second minimum $\Delta g_{2,3}^{max} = g_3^{max} - g_2^{max}$ improves from -0.20 in PBE to +0.18 in SCAN, leading to a significant improvement towards the experimental value of +0.4. However, SCAN still gives an over-structured $g_{OH}(r)$ and the discrepancy is larger than that of $g_{OO}(r)$. We rationalize this discrepancy based on the fact that NQEs affect H atoms more severely than O atoms [8,9]; an explicit treatment of NQEs would be necessary to further examine the $g_{OH}(r)$.

After discussions of the $g_{OO}(r)$ and $g_{OH}(r)$, we discuss the H-H radial distribution function, $g_{HH}(r)$, as illustrated in Fig. 3. As listed in Table III, for the first maximum of the $g_{HH}(r)$, we observe that while PBE and PBE+vdW give nearly identical results, SCAN yields a higher first maximum $g_1^{max}$ and a decreased position $r_1^{max}$. As the first peak corresponds to the distance between the two H atoms within a water molecule, these changes support that SCAN alters the intramolecular short-range interactions and enhances the covalency of water molecules. Meanwhile, the improvements over the second maximum and the second minimum by PBE+vdW and SCAN compared to PBE are similar to the previously discussed cases of first maximum of the $g_{OO}(r)$ and the second maximum of the $g_{OH}(r)$. These improvements indicate that SCAN corrects the water structure more significantly than PBE+vdW does because the former functional not only captures the intermediate-range vdW interactions but also reduces the directional H-bond strength.

## B. Oxygen-oxygen-oxygen triplet angular distribution function and tetrahedrality

In order to further analyze the local arrangement of water molecules in liquid phase, we have computed the distribution of triplet oxygen-oxygen-oxygen angles, $P_{OOO}(\theta)$, within the first coordination shell (see Fig. 4) and the tetrahedral order parameter $q$ (see Table IV). To compute the $P_{OOO}(\theta)$, three O atoms were considered as part of a given triplet if two of the O atoms were within a cutoff distance from the third one, so that the average O-O coordination number reaches 4.0 at the cutoff distance [10,28]. The cutoff distances were selected as 3.26, 3.24, and 3.20 Å from PBE, PBE+vdW, and SCAN AIMD simulations, respectively. The tetrahedral order parameter $q$ is determined by

$$q = 1 - \frac{3}{8}\sum_{i=1}^{3}\sum_{j=i+1}^{4}(\cos(\theta_{ij}) + \frac{1}{3})^2 \ , \tag{1}$$

and varies between 0 and 1, where 0 represents an ideal gas while 1 represents a perfect tetrahedral structure. The computed results of $P_{OOO}(\theta)$ and $q$ from our AIMD simulations are also compared with results of EPSR based on the joint X-ray/neutron data [10].

It can be inferred from Fig. 4 that the EPSR-experimental $P_{OOO}(\theta)$ exhibits a broad distribution with its maximum at around 100.5°, suggesting that the local tetrahedral network in liquid water is considerably more disordered than that in ice. At the PBE level of theory, the overall distribution of $P_{OOO}(\theta)$ is significantly narrower compared to the EPSR-experimental result. Similarly, the tetrahedral order parameter $q$ in our PBE simulation has a value of 0.895, which is much higher than the EPSR-experimental value of 0.576 [10]. Both $P_{OOO}(\theta)$ and $q$ computed from the PBE-based AIMD simulation corroborates with the aforementioned fact that liquid water modeled by PBE functional tends to be over-structured with overly strong H-bonding. In Ref. [28], the simulation of liquid water by PBE at 300 K gives a $P_{OOO}(\theta)$ distribution with the peak at ~109°, which is very close to the perfect tetrahedral angle of 109.5°. In our simulation, the peak position is shifted to 104° at an elevated temperature of 330 K. We consider that the ~5° difference of the peak position originates from the additional 30 K in our simulation, as when temperature increases, the topological structure close to the perfect tetrahedral network is less favored. Noticeably, two simulations utilizing PBE0+vdW level of theory in Ref. [28] at 300 and 330 K gave rise to peak positions of $P_{OOO}(\theta)$ at around 106° and 102°, respectively, the ~ 4° difference caused by elevated temperature is in consistency with our simulation results.

The degree of tetrahedral order in liquid water is slightly reduced when vdW interactions are included in the simulation. It is observed that the distribution of $P_{OOO}(\theta)$ as calculated from the PBE+vdW trajectory is widened with a noticeable reduction of peak intensity when compared to that obtained from the PBE trajectory, while the tetrahedral order parameter $q$ reduces from 0.895 (PBE) to 0.839 (PBE+vdW). SCAN further improves the tetrahedral order as evidenced by the fact that the peak intensity is lowered, the distribution of $P_{OOO}(\theta)$ is widened, and the $q$ value is reduced to 0.709. Also, we find that the peak position of $P_{OOO}(\theta)$ keeps the same value of 104° by both PBE and PBE+vdW but reduces to 101° by SCAN, which is very close to the experimental value of 100.5°. This is rationalized by the fact that the peak position is correlated with the most perfect local tetrahedral structure that formed by directional H-bonds, which are weakened by SCAN as previously mentioned. The results shown here suggest that SCAN also accurately captures the local tetrahedral structure of liquid water.

### C. Hydrogen-bond analysis

In an ice system, generally, each water molecule stably donates two and accepts two H-bonds, resulting in four H-bonds (intact H-bonds) on average. The fluidity of liquid water comes from the fact that H-bonds, which form the tetrahedral-like network, are continuously breaking and forming on a timescale of ps. By comparing the statistics of H-bonds in different AIMD trajectories, we obtain information on the H-bond strength and the fluidity of the system. The notion of H-bond itself, however, is not uniquely defined; although any quantitative description of the statistics for intact H-bonds yields certain degree of ambiguity, a general agreement among many proposed definitions for intact H-bonds still exist [62]. We follow the H-bond definition proposed by Luzar and Chandler [63] and define an intact H-bond when O-O distance $R_{OO} < 3.5$ Å and $\beta <$ 30°, where $\beta \equiv \angle O_A \cdots O_D - H_D$ denotes the bond angle formed by an oxygen atom $O_A$ that accepts a H-bond ($H_D$) from a nearby oxygen atom $O_D$.

Using this definition of H-bonds, we compute the average numbers of H-bonds per water molecule in different AIMD trajectories (see Table IV), which are 3.85, 3.77, and 3.60 for PBE, PBE+vdW, and SCAN functionals, respectively. We observe that while both PBE+vdW and SCAN functionals increase the proportion of broken H-bonds and the degree of disorder in the local tetrahedral network of liquid water, the former functional has a smaller effect than the latter one. As discussed in Secs III A and III B, vdW interactions disturb the H-bond network by populating the interstitial region and disrupting the local tetrahedral structure. SCAN also captures the intermediate-range vdW interactions but additionally improves the covalency of water molecules and electronic structures (see Sec. III E). Moreover, SCAN weakens the overly strong directional H-bonding strength described by PBE, which vdW interactions fail to correct. The improved description of directional H-bonding by SCAN contributes to the further reduction of average number of H-bonds when compared to PBE+vdW.

To further illustrate the H-bond statistics in liquid water, we compute the time-averaged distribution of H-bonds in terms of their numbers (see Fig. 5). In the PBE trajectory, the proportion of water molecules that have four H-bonds is 79.5%; this number reduces to 69.7% and 56.4% in the PBE+vdW and SCAN trajectories, respectively. Therefore, we find that both PBE+vdW and SCAN functionals reduce the proportion of water molecules with four H-bonds and increase the proportions of water molecules with 1, 2, 3, and 5 H-bonds. Once again, SCAN functional affects the H-bond statistics more than PBE+vdW does. This result is consistent with our previous analyses that the PBE+vdW and SCAN functionals influence the H-bond network of liquid water differently.

The improvement over the average number of H-bonds from our AIMD simulations also implies an enhancement of the fluidity of liquid water, since the breakage and formation of H-bonds through thermal fluctuations affect diffusivity of water molecules: as there are less H-bonds with the H-bonding strength reduced, the diffusion coefficient $D$ increases (see Table IV). These $D$ values are extracted from the slope of the molecular mean square displacement (MSD) versus the simulation time. However, these $D$ values can be further improved based on the following considerations. First, small periodic cells have a non-negligible finite-size effect on $D$ [26,64], in which diffusion coefficients converge slowly with the increased sizes of the simulation cell. Second, a much longer trajectory or multiple runs can reduce the uncertainty of $D$ [65].

**D. Ring analyses**

We can gain more understanding of the liquid water structure by analyzing the ring statistics. The distribution of closed rings were used to study topological networks in silicate network structures [66,67] but also in liquid water systems [68-70], where rings are threaded by O atoms that are connected with H-bonds. In this work, rings were defined adopting the shortest-path definition from Ref. [68], in which only the shortest circuit with two continuous H-bonds from one water molecule is recorded. The analyses of rings include average size of rings, distribution of ring sizes, average number of rings, and distribution of ring numbers (see Fig. 6 and Table IV). Note that a ring size is defined as the number of O atoms in a closed ring.

In an ice I$h$ system, there are in general 12 6-membered rings threading through each water molecule. In other words, a water molecule is involved in 12 rings, with each ring threading 6 water molecules. However, in a liquid water system, one would expect different ring sizes and ring numbers for different sizes, as the broken H-bonds and disordered H-bonding network give rise to rings with irregular shapes and sizes other than 6. We define small rings as those threading 4 molecules and less, medium rings as those threading 5 to 8 molecules, and large rings as those threading more than 8 molecules, by loosely choosing the cross points in the probability distributions of ring sizes by different XC functionals in Fig. 6(a). The existence of small and medium rings is mainly caused by the disruption of local tetrahedrality in liquid water, while the large rings are mainly formed by the combination of two or more small and/or medium rings when shared H-bond(s) between adjacent rings break by thermal fluctuations. As shown in Fig. 6(a), both PBE+vdW and SCAN reduce the numbers of small and medium rings and increase the numbers of large rings, with SCAN outperforms PBE+vdW. This result is consistent with the aforementioned structural analyses and further confirms our observation that SCAN describes a more disordered H-bonding network in liquid water.

We further examine the statistics for the sizes of rings as shown in Fig. 6(a). Both PBE+vdW and SCAN functionals reduce the portion of medium rings and increase the portions of small and large rings as compared to the PBE results. Particularly, the proportion of 6-membered rings, which is the dominating size of rings found in liquid water in all the three functionals tested, has been reduced by PBE+vdW as compared to the PBE result; the proportion of 6-membered ring is further reduced by SCAN. Notably, SCAN significantly increases the portion of rings threading more than 8 water molecules. For instance, the average numbers of 10-membered rings obtained by PBE, PBE+vdW, and SCAN functionals are 0.12, 0.21, 0.60, respectively. Accordingly, the average ring sizes for PBE, PBE+vdW, and SCAN functionals are 6.64, 7.01, and 7.75, respectively. The reduction of medium rings and increase of small and large rings in both PBE+vdW and SCAN functionals are associated with the fact that both functionals disrupt the local tetrahedrality, soften the coordination shells, and increase the portion of broken H-bonds (see Secs. III B and III C). However, the influence on the distribution of ring sizes by SCAN is more significant than PBE+vdW because SCAN not only captures the intermediate-range vdW interactions but also corrects the overly strong directional H-bond strength in the PBE simulation.

Fig. 6(b) shows the distributions of number of rings threading one water molecule, in which we find that the distribution is broadened with more small and large numbers of rings by PBE+vdW as compared to the one by PBE, and is even more broadened when SCAN is considered. A broader distribution reflects a H-bond network that is softer and more distorted, which is consistent with the fact that SCAN has a larger effect on disrupting the H-bond network.

## E. Electronic properties

In this section, we discuss electronic properties of liquid water which govern the strength of H-bonds and influence solvation behaviors. We first analyze the electronic density of states (DOS) and band gaps of liquid water obtained by AIMD simulations with different XC functionals, and then discuss molecular dipole moments. As we will see, the analyses on electronic properties show that the SCAN functional improves the electronic structure of molecules in liquid water, which leads to the correction of the overly strong H-bonds by PBE, while PBE+vdW does not provide such improvement.

The improvement in the electronic properties of liquid water achieved by the SCAN functional can be seen through the comparison of the computed electronic DOS with the full valence band photoemission spectroscopy [15] (see Fig. 7). The DOS of liquid water in each AIMD simulation was computed through a Gaussian broadening of 0.5 eV [71] for eigenstates of the system and averaged over 10 snapshots through the trajectory. The four peaks of the DOS are assigned to the $2a_1$, $1b_2$, $3a_1$, and $1b_1$ valence molecular orbitals based on the spatial symmetries of a water molecule, with the lone electron pairs closely connected to the $2a_1$ and $1b_1$ orbitals, and the bonding electron pairs related to the $1b_2$ and $3a_1$ orbitals. The simulated DOS and the photoemission spectra are aligned following Ref. [72] in such a way that the onset position, which was taken as the $x$ intercept of the linear fitting for the tail of the $1b_1$ distribution, is set to 0 eV.

From Fig. 7, we notice that PBE+vdW yields an almost identical DOS compared to PBE, which is in accordance with the previous report that the inclusion of vdW interactions do not lead to any significant effect on either DOS or the band gap [73], whereas the electronic structure is improved by SCAN. We find that in SCAN, the $2a_1$ peak shifts 1.38 eV to a smaller energy compared to PBE, which is consistent with previous study [54] and agrees better with the experimental value. As a result, the energy differences $\Delta E$ between the $2a_1$ and $1b_1$ peaks predicted by PBE, PBE+vdW and SCAN are 17.63, 17.63, and 18.98 eV, respectively, as compared to the experimental value of 19.74 eV [15]. This suggests that compared to the value obtained by PBE, SCAN significantly improves the energy difference between the two states and better describes the lone pair electrons. We have also computed the electronic band gaps as obtained through an average of 10 configurations for each functional. The band gaps are $4.51 \pm 0.10$, $4.32 \pm 0.14$, and $4.90 \pm 0.08$ eV in AIMD simulations utilizing PBE, PBE+vdW, and SCAN functionals, respectively. The band gaps obtained with the GGA level of theory are consistent with previous reports [6], while the band gap calculated with SCAN is corrected towards, although still much lower than, the experimental value of 8.7 eV [74].

Next, we examine the distribution of the molecular dipole moments in liquid water. For reference, the dipole moment of a single water molecule in the gas phase is accurately known to be 1.855 D from electroscopic experiment [75], and several DFT functionals are capable of reproducing this experimental value to within 3% [6]. However, the molecular dipole moment of liquid water extracted from X-ray scattering form factors yields a large uncertainty ($2.9 \pm 0.6$ D) [76]. In the meantime, for the calculation of dipole moments from AIMD simulations of condensed-phased systems, the partitioning of the electron density would be required and different partitioning schemes would lead to widely different values of dipole moments [77]. In this regard, maximally localized Wannier functions (MLWFs), which are obtained through a unitary transformation of occupied Kohn-Sham orbitals, have been shown to be very useful in computing dipole moments in condensed-phase environments [6,78-80]. It has been shown that in bulk water the amount of overlap between the charge distribution of Wannier functions associated with different molecules is less than 1% of the norm [81], which allows for a nearly unique definition of the charges belonging to a given water molecule and eliminates to a large extent the partitioning issues when computing dipole moments in liquid water.

In our simulation, four doubly occupied valence MLWFs represent centers of eight valence electrons associated with each water molecule. Two of the MLWFs centered on the O-H covalent bonds are defined as bonding pair electrons, while the other two being approximately centered on the remaining tetrahedral directions are labelled as lone pair electrons. The dipole moment ($\boldsymbol{\mu}$) for a water molecule can be computed as follows:

$$\boldsymbol{\mu} = \mathbf{R}_{H_1} + \mathbf{R}_{H_2} + 6\mathbf{R}_O - 2\sum_{i=1}^{4}\mathbf{R}_{W_i}, \tag{2}$$

in which $\mathbf{R}_{H_1}$, $\mathbf{R}_{H_2}$, and $\mathbf{R}_O$ are the Cartesian coordinates of the two H atoms and one O atom of a water molecule, respectively, and $\mathbf{R}_{W_i}$ are the coordinates of the four corresponding MLWF centers. With this formula, we have computed the distribution of dipole moments in liquid water from AIMD trajectories with PBE, PBE+vdW, and SCAN XC functionals and the results are shown in Fig. 8(a). It is evident that the peak positions of the dipole distributions in Fig. 8(a) obtained from PBE+vdW and SCAN functionals are shifted towards smaller values as compared to that from PBE. In particular, the shift in the dipole maximum is much larger with SCAN than PBE+vdW. The average dipole moments are 3.19, 3.13, and 2.94 D for simulations with PBE, PBE+vdW, SCAN functionals, respectively. Additionally, we note that the height of the peak for the distribution of dipole moments from the PBE+vdW simulation is the lowest among the three AIMD simulations, which is associated with the widening of distribution of lone pair O-MLWF distances, as will be explained next.

To further understand the variation of dipole moments in liquid water given by different XC functionals, we also illustrate in Fig. 8(b) the electronic properties of liquid water by computing the distribution of distance between an O atom of water molecule and the centers of its four nearest MLWFs ($R_{O-MLWF} = |\mathbf{R}_O - \mathbf{R}_W|$). In general, Fig. 8(b) shows that the two bonding pair electrons have a distribution of distances longer than $R_{O-MLWF} = 0.45$ Å, while the two lone pair electrons are well separated from the bonding pair electrons and have a distribution of distances shorter than $R_{O-MLWF} = 0.40$ Å. We also list the O-MLWF distances for a single water molecule for reference. Specifically, the PBE, PBE+vdW, and SCAN functionals respectively yield 0.523, 0.523, and 0.518 Å as the distances between bonding pair electrons and the oxygen. Moreover, the distances between lone pair electrons and the oxygen are 0.289, 0.289, and 0.290 Å as obtained from the PBE,

PBE+vdW, and SCAN functionals, respectively. In the liquid phase, a water molecule accepts an H-bond through the electrostatic interaction between one of its lone pair electrons and an H atom from one of its neighboring water molecules. If both the H atom and the lone pair electron are closer to the O atom, then the accepting side of the H-bond, i.e., the O atom is in a less negative electric environment. Meanwhile, the donating side of the H-bond, i.e., the H atom is in a less positive electric environment.

Notably, the distributions of bonding pair electrons by the three XC functionals only exhibit small differences, while the PBE+vdW and SCAN functionals change more significantly the distribution of the lone pair electrons. Specifically, the PBE+vdW functional decreases the peak position of the lone pair electrons from 0.338 Å, as obtained by the PBE functional, to 0.334 Å, and lowers the peak intensity and broadens the distribution. As discussed before, smaller distances between O atoms and their corresponding lone pair electrons lead to less negative environments for the O atoms, which in turn accept less H-bonds. This is consistent with the fact that single molecules in gas phase have no H-bonds and relatively smaller electric dipole moments as compared to those in condensed phases. Therefore, these changes of dipole moments and centers of MLWFs can be rationalized by the inclusion of vdW interactions which populate the interstitial region between the first and second solvation shells and reduce the average number of H-bonds in liquid water.

The SCAN functional yields a distribution of lone pair electrons with almost unchanged broadness, but the peak position shifts 0.012 Å inwards to 0.326 Å as compared to that from PBE. The appreciable net reduction of the $R_{O-MLWF}$ distance and covalent bond length (see Sec. III A) by SCAN contribute as the main sources to the decrease of the average dipole moment for liquid water. Importantly, SCAN changes covalent bond strengths in liquid water which results in the reduced polarizability and the weakened H-bond strength; this effect is absent in the modeling of liquid water using PBE+vdW. In conclusion, this unique change of electrostatic interactions by SCAN functional provides a more accurate description of H-bond network in liquid water, and is consistent with previous analyses on the structural and dynamical properties of liquid water by SCAN functional that differ from the simulation with the PBE+vdW functional.

## F. NVT and the NpT ensembles

In this section, we compare results obtained from simulations using different ensembles (NVT and NpT) with PBE and SCAN functionals for ambient liquid water. We have taken the NpT ensemble results from the previous work by Chen etc. [54] and compared with our current NVT simulations. In principle, these two ensembles should yield the same structure of liquid water if the underlying potential is sufficiently accurate to predict the correct equilibrium volume (V) at a given external pressure (p) and the simulation time is long enough for sampling the phase space. However, the equilibrium volumes/densities at 1 bar predicted with both PBE and SCAN functionals in the NpT ensemble deviate from the experimental value (1 g/mL). As a result, we find in Fig. 9 that the NVT ensemble yields a slightly more structured liquid water (higher first and second peaks in $g_{OO}(r)$) when compared to that from the NpT ensemble.

According to the recent study by Chen *et al.*, the average densities of liquid water obtained from AIMD simulations with PBE and SCAN functionals in the NpT ensemble were 0.850 and 1.050 g/mL, respectively [54]; another work on the simulation of liquid water also shows that utilizing SCAN overcorrects the density of water compared to experiment [82]. On the other hand, the density of liquid water is fixed as 1.0 g/mL in our NVT simulation. Therefore, the smaller density as predicted by the PBE AIMD trajectory in the NpT ensemble implies that water molecules are more separated than those in the SCAN AIMD trajectory. Indeed, we observe that with the PBE functional, the positions of maximums and minimums of $g_{OO}(r)$ in the NpT ensemble shift outwards compared to those in the NVT ensemble. Specifically, the position of the first maximum $r_1^{max}$ increases from 2.71 Å in the NVT ensemble to 2.73 Å in the NpT ensemble. Accordingly, the averaged pressures of liquid water, calculated from 50 snapshots from the NVT ensemble with the kinetic energy cutoff set as 150 Ry, are 0.9±3.7, -7.0±4.1, and -6.5±3.4 kbar in the trajectories adopting the PBE, PBE+vdW, and SCAN functionals, respectively. By softening the structure of liquid water, SCAN corrects the large density error in PBE in the NpT ensemble. However, the use of the experimental density in the NVT ensemble (1.0 g/mL) than the one predicted by NpT (1.05 g/mL) mitigates the correction effect by SCAN. As a result, water molecules in the NVT ensemble are more apart with their shell structures become slightly over-structured; we find that the positions of maximums and minimums in $g_{OO}(r)$ as obtained from the SCAN trajectory in the NpT ensemble are squeezed slightly inwards compared to those from the NVT ensemble, which is consistent with the slightly higher density obtained in the NpT ensemble. To be specific, the position of the first maximum $r_1^{max}$ decreases from 2.75 Å in the NVT to 2.74 Å in the NpT. Consistently, the self-diffusion coefficient of 0.129 Å²/ps in liquid water obtained from the SCAN trajectory in the NVT ensemble is smaller than the value of 0.190 Å²/ps from the NpT ensemble.

## IV. CONCLUSIONS

We have performed AIMD simulations of liquid water at ambient conditions in the canonical ensemble with the focus on comparing the results as obtained from the newly developed meta-GGA SCAN XC functional to those calculated with the PBE and PBE+vdW functionals. Importantly, SCAN captures the intermediate-range vdW interaction that is crucial in improving the structure of liquid water. Besides, SCAN provides a better description of electronic structures that lead to a more delicate balance between covalent bonds and H-bonds in liquid water, largely curing the overly-strong H-bonds and softening the rigid solvation shell structure as prescribed by PBE functional. We find that SCAN yields more accurate structural properties like radial distribution functions and O-O-O triplet angular distribution functions, improves the H-bond network as can be shown in the H-bond statistics and ring statistics, and predicts electronic properties including DOS, band gap, and dipole moment that are closer to experimental data. Furthermore, we discussed properties of liquid water by SCAN in both NVT and NpT ensembles. In general, SCAN performs better than both PBE and PBE+vdW functionals for describing liquid water. However, there are still improvements to be made regarding the structure of liquid water. For instance, the current modeling by SCAN still prescribes a slightly over-structured solvation shell structures. The correction for SIE by hybrid functional was reported to further mitigate the over-structuring [28,43] but the performance of combining SCAN with the hybrid functional is still unknown and can be investigated in future works. Additionally, the explicit treatment of NQEs through the Feynman discretized path-integral approach [8,9,72] can be used to correct the discrepancies regarding the widths and intensities of radial distribution functions, especially of $g_{OH}(r)$. Nevertheless, the SCAN-based AIMD simulation in the canonical ensemble already describes a delicate balance among covalent bonds, H-bonds, and vdW interactions that lead to reasonably accurate structures and dynamics of liquid water. Therefore, the SCAN XC functional provides a promising *ab initio* model of liquid water and should be considered for future work on modeling of aqueous solutions.

**ACKNOWLEDGEMENTS**

This work was supported by National Science Foundation through Awards DMR-1552287 (X.W.). This research used resources of the National Energy Research Scientific Computing Center, which is supported by the U.S. Department of Energy (DOE), Office of Science under Contract DEAC0205CH11231. The work of Z.S. was supported as part of the Center for the Computational Design of Functional Layered Materials, an Energy Frontier Research Center funded by the U.S. DOE, Office of Science, Basic Energy Sciences under Award DESC0012575. The work of L. Z. was partially supported by the American Chemical Society Petroleum Research Fund (ACS PRF) under Grant No. 53482-DNI6.

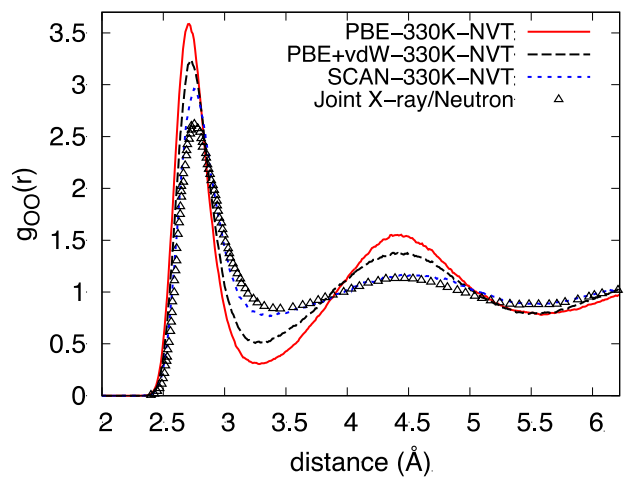

Figure 1: (Color online) O-O radial distribution functions $g_{OO}(r)$ of liquid water as obtained from AIMD simulations in the canonical ensemble employing PBE (red solid line), PBE+vdW (black dashed line), and SCAN exchange-correlation functionals (blue dashed line). The experimental data are from joint X-ray/neutron experiments [10].

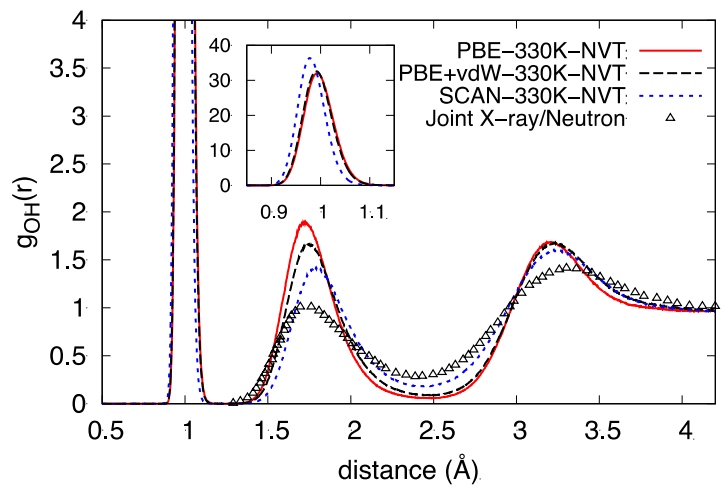

Figure 2: (Color online) O-H radial distribution functions $g_{OH}(r)$ of liquid water as obtained from AIMD simulations in the canonical ensemble employing PBE (red solid line), PBE+vdW (black dashed line), and SCAN exchange-correlation functionals (blue dashed line). The experimental data are from joint X-ray/neutron experiments [10].

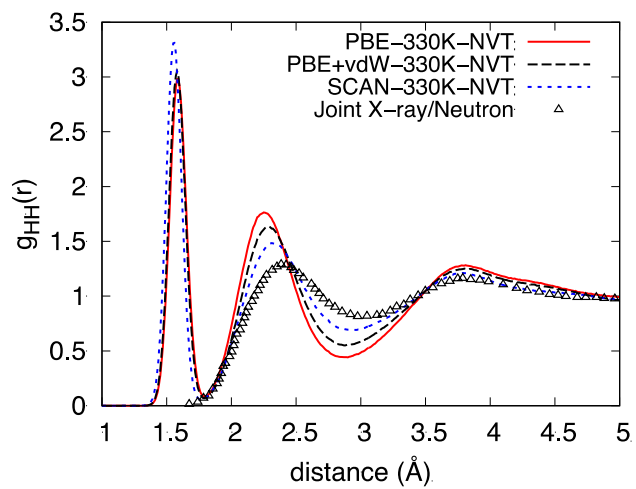

Figure 3: (Color online) H-H radial distribution functions $g_{HH}(r)$ of liquid water as obtained from AIMD simulations in the canonical ensemble employing PBE (red solid line), PBE+vdW (black dashed line), and SCAN XC functionals (blue dashed line). The experimental data are from joint X-ray/neutron experiments [10].

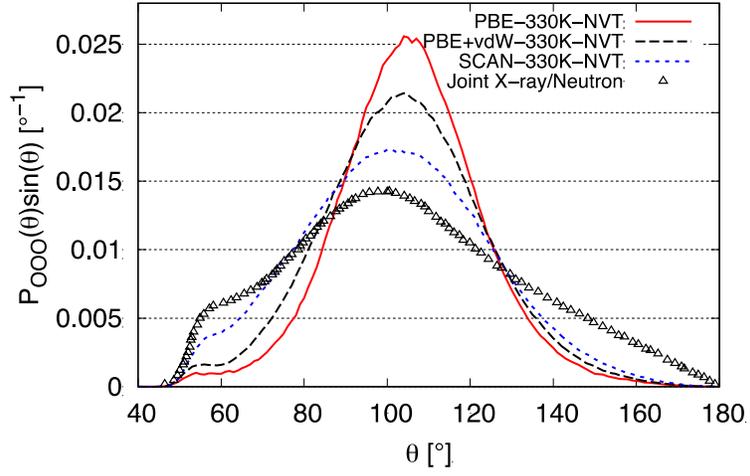

Figure 4: (Color online) O-O-O triplet angular distribution functions $P_{OOO}(\theta)$ of liquid water obtained from AIMD simulations and EPSR based on joint X-ray/neutron scattering data [10]. The triplet angular distribution functions show here were normalized to $\int_0^\pi d\theta P_{OOO}(\theta) \sin(\theta) = 1$.

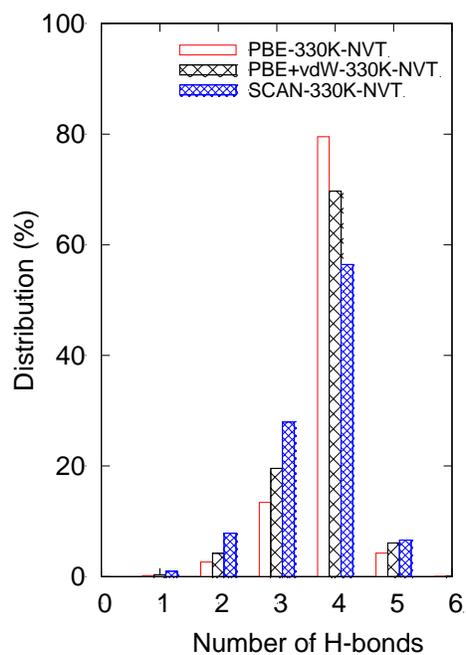

Figure 5: (Color online) Distribution of number of H-bonds in liquid water obtained from AIMD simulations employing PBE, PBE+vdW, and SCAN XC functionals.

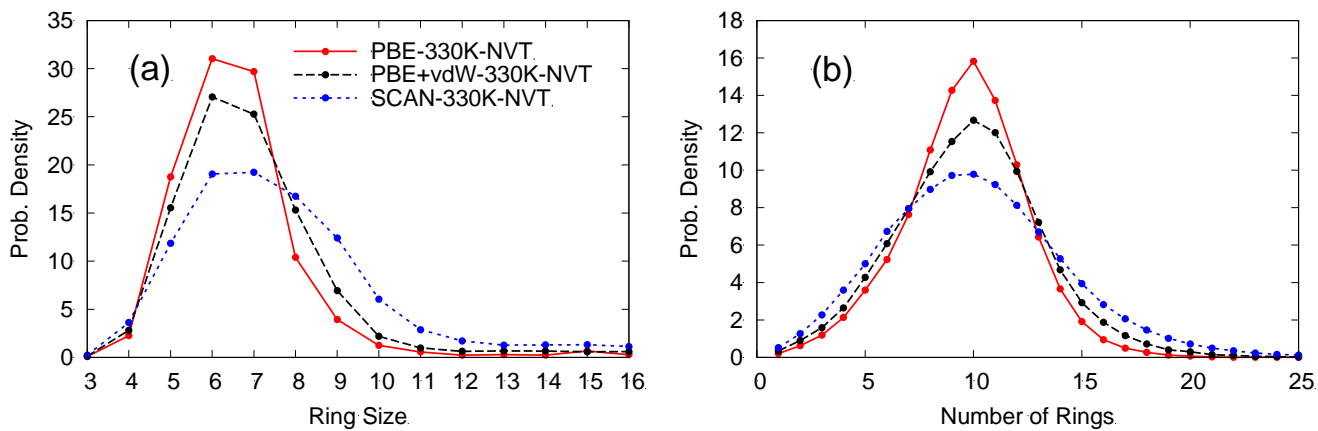

Figure 6: (Color online) (a) Probability distribution of rings as a function of ring sizes obtained from AIMD simulations by using PBE, PBE+vdW, and SCAN functionals. The integration of the area for each curve is the average number of rings threading one water molecule. (b) Distribution of number of rings threaded through one water molecule obtained from AIMD simulations by employing PBE, PBE+vdW, and SCAN functionals.

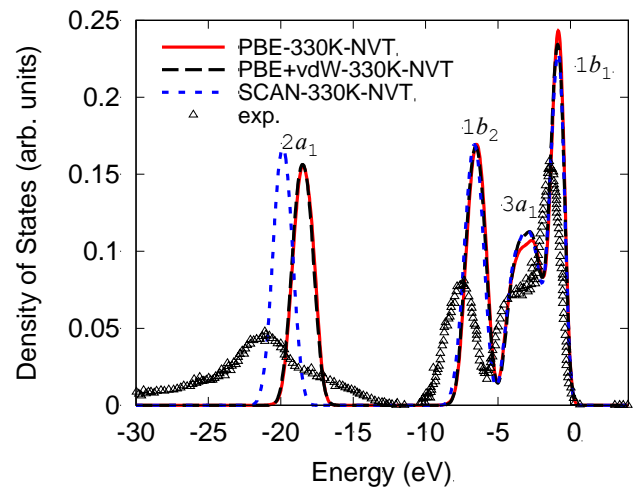

Figure 7. (Color online) Density of states (DOS) of water obtained from AIMD simulations employing PBE, PBE+vdW, and SCAN XC functionals. The experimental data of DOS from photoemission spectra [15] are also shown.

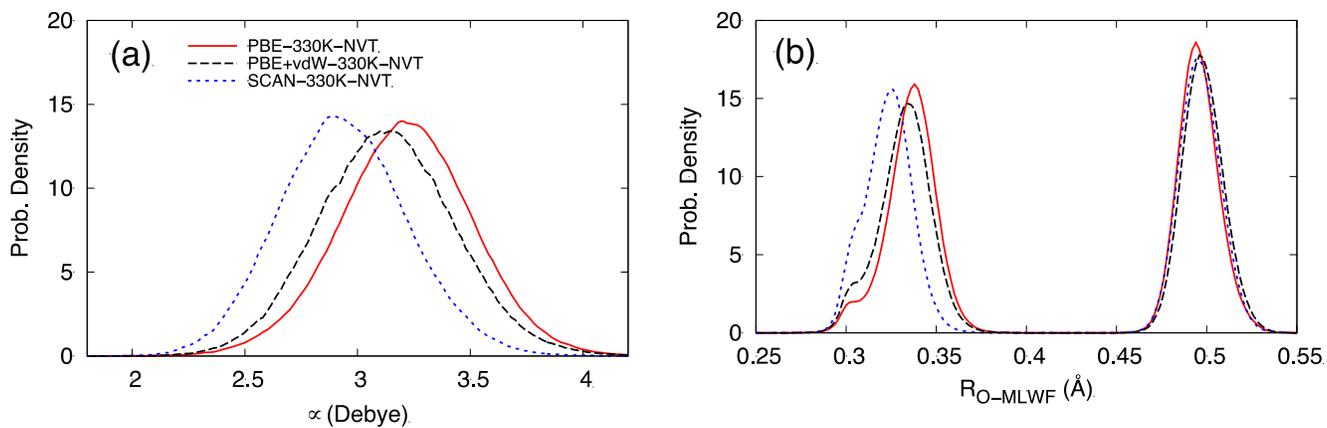

Figure 8: (Color online) (a) Probability density plots of the distributions of molecular dipole moment magnitudes ($\mu$ in Debye) in liquid water obtained from the AIMD simulations employing PBE, PBE+vdW, and SCAN XC functionals. (b) Probability density plots of the distributions of distance between the oxygen atoms in a given water molecule and the centers of the four nearest maximally localized Wannier functions ($R_{\text{O-MLWF}}$ in Å).

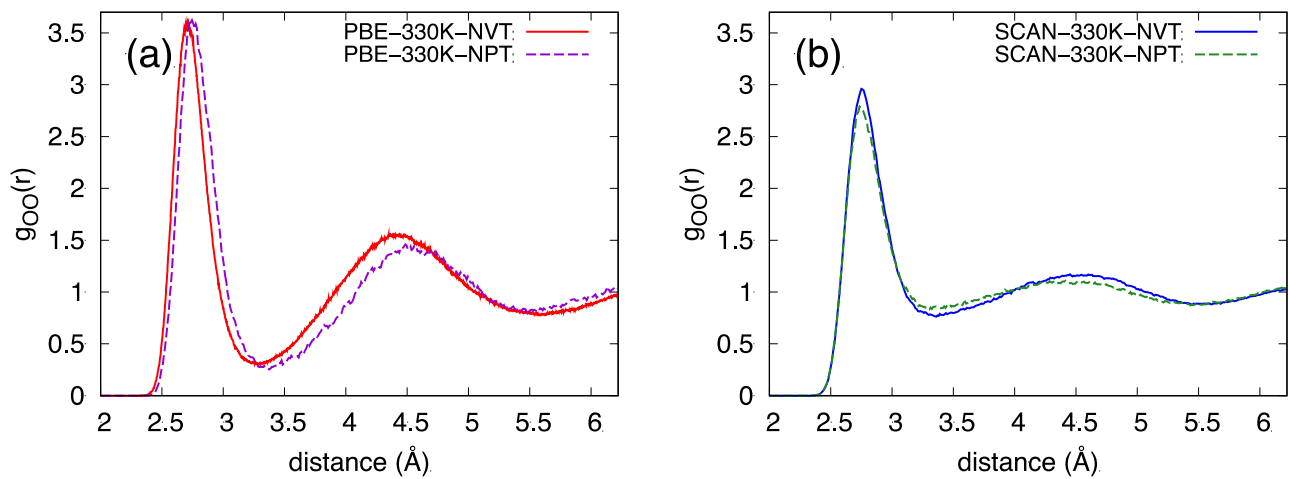

Figure 9: (Color online) (a) O-O radial distribution functions $g_{OO}(r)$ of liquid water obtained from PBE-based AIMD simulations in the NVT (red solid line) and the NpT (purple dashed line) ensembles. (b) O-O radial distribution functions $g_{OO}(r)$ of liquid water obtained from SCAN-based AIMD simulations in the NVT (blue solid line) and the NpT (green dashed line) ensembles.

Table I: Tabulated summaries of the structural properties of liquid water from the O-O radial distribution functions $g_{OO}(r)$ obtained from AIMD simulations in the canonical ensemble and experimental data. Three different XC functionals, namely, PBE, PBE+vdW, and SCAN are used in AIMD simulations. From left to right, positions (in Å) and intensities of the first maximum ($r_1^{max}$ and $g_1^{max}$), first minimum ($r_1^{min}$ and $g_1^{min}$), and second maximum ($r_2^{max}$ and $g_2^{max}$) of the corresponding $g_{OO}(r)$.

| $g_{OO}(r)$ | $r_1^{max}$ | $g_1^{max}$ | $r_1^{min}$ | $g_1^{min}$ | $r_2^{max}$ | $g_2^{max}$ |
|---|---|---|---|---|---|---|
| PBE | 2.71 | 3.59 | 3.28 | 0.31 | 4.44 | 1.55 |
| PBE+vdW | 2.72 | 3.24 | 3.25 | 0.51 | 4.38 | 1.38 |
| SCAN | 2.75 | 2.96 | 3.36 | 0.76 | 4.47 | 1.15 |
| Expt. [17] | 2.80 | 2.57 | 3.45 | 0.84 | 4.50 | 1.12 |
| Expt. [10] | 2.75 | 2.62 | 3.45 | 0.84 | 4.43 | 1.13 |

Table II: Tabulated summaries of the structural properties of liquid water from the O-H radial distribution functions $g_{OH}(r)$ obtained from AIMD simulations in the canonical ensemble and experimental data. From left to right, positions (in Å) and intensities of the first maximum ($r_1^{max}$ and $g_1^{max}$), second maximum ($r_2^{max}$ and $g_2^{max}$), second minimum ($r_2^{min}$ and $g_2^{min}$), and third maximum ($r_3^{max}$ and $g_3^{max}$) of the corresponding $g_{OH}(r)$.

| $g_{OH}(r)$ | $r_1^{max}$ | $g_1^{max}$ | $r_2^{max}$ | $g_2^{max}$ | $r_2^{min}$ | $g_2^{min}$ | $r_3^{max}$ | $g_3^{max}$ |
|---|---|---|---|---|---|---|---|---|
| PBE | 0.994 | 32.1 | 1.72 | 1.88 | 2.49 | 0.06 | 3.21 | 1.68 |
| PBE+vdW | 0.990 | 32.6 | 1.75 | 1.66 | 2.45 | 0.09 | 3.23 | 1.67 |
| SCAN | 0.978 | 36.4 | 1.79 | 1.42 | 2.40 | 0.18 | 3.25 | 1.60 |
| Expt. [17] | - | - | 1.73 | 1.01 | 2.39 | 0.28 | 3.30 | 1.41 |

Table III: Tabulated summaries of the structural properties of liquid water from the H-H radial distribution functions $g_{HH}(r)$ obtained from AIMD simulations in the canonical ensemble and experimental data. From left to right, positions (in Å) and intensities of the first maximum ($r_1^{max}$ and $g_1^{max}$), second maximum ($r_2^{max}$ and $g_2^{max}$), second minimum ($r_2^{min}$ and $g_2^{min}$), and third maximum ($r_3^{max}$ and $g_3^{max}$) of the corresponding $g_{HH}(r)$.

| $g_{HH}(r)$ | $r_1^{max}$ | $g_1^{max}$ | $r_1^{min}$ | $g_1^{min}$ | $r_2^{max}$ | $g_2^{max}$ | $r_2^{min}$ | $g_2^{min}$ |
|---|---|---|---|---|---|---|---|---|
| PBE | 1.59 | 2.98 | 1.79 | 0.08 | 2.25 | 1.76 | 2.90 | 0.44 |
| PBE+vdW | 1.58 | 3.05 | 1.79 | 0.07 | 2.29 | 1.63 | 2.87 | 0.55 |
| SCAN | 1.56 | 3.31 | 1.79 | 0.05 | 2.31 | 1.48 | 2.93 | 0.69 |
| Expt. [10] | - | - | - | - | 2.38 | 1.29 | 2.98 | 0.82 |

Table IV: Tabulated summary of the properties of liquid water obtained via the AIMD simulations performed in this work and various experiments. From left to right, the average number of H-bonds ($n_{HB}$) per water molecule; the average tetrahedrality parameter ($q$) per water molecule; the average size of the rings ($S_{ring}$) threaded through one water molecule; the average number of the rings ($N_{ring}$) threaded through one water molecule; the average diffusion coefficient ($D$ in Å$^2$/ps) per AIMD simulation; the average dipole moment ($\mu$ in Debye) per water molecule; the energy difference between the peaks of $2a_1$ and $1b_1$ valence molecular orbits ($\Delta E = E_{1b_1} - E_{2a_1}$ in eV) in DOS; the electronic band gaps ($E_g$) from each AIMD simulations (in eV). For reference, the available corresponding experimental data are provided in the last row.

| Method | $n_{HB}$ | $q$ | $S_{ring}$ | $N_{ring}$ | $D$ | $\mu$ | $\Delta E$ | $E_g$ |
|--------|----------|-----|-----------|-----------|-----|-------|-----------|-------|
| PBE | $3.85 \pm 0.52$ | $0.895 \pm 0.459$ | $6.64 \pm 1.69$ | $9.63 \pm 2.84$ | $0.015 \pm 0.012$ | $3.19 \pm 0.29$ | 17.63 | $4.51 \pm 0.10$ |
| PBE+vdW | $3.77 \pm 0.64$ | $0.839 \pm 0.478$ | $7.01 \pm 2.04$ | $9.81 \pm 3.39$ | $0.041 \pm 0.023$ | $3.13 \pm 0.30$ | 15.63 | $4.32 \pm 0.14$ |
| SCAN | $3.60 \pm 0.77$ | $0.709 \pm 0.473$ | $7.75 \pm 2.58$ | $10.01 \pm 4.22$ | $0.129 \pm 0.046$ | $2.94 \pm 0.28$ | 18.98 | $4.90 \pm 0.08$ |
| Expt. | 3.58 [83] | 0.576 [10] | - | - | 0.186 [84] | 2.9 [76] | 19.74 [15] | 8.7±0.6 [72] |